\documentclass[%
superscriptaddress,
twocolumn,
10pt,
 amsmath,amssymb,
 aps,
 pra,
floatfix]{revtex4-2}

\usepackage{graphicx,array}
\usepackage{dcolumn}
\usepackage{multirow}
\usepackage{array}
\newcolumntype{P}[1]{>{\centering\arraybackslash}p{#1}}
\usepackage{enumitem}
\usepackage{hyperref}
\usepackage{longtable}
\usepackage{xcolor}

\begin{document}
\title{Network analysis approach to Likert-style surveys}

\author{Robert P. Dalka}
\email{rpdalka@umd.edu}
\affiliation{University of Maryland, College Park, Maryland 20742, USA}

\author{Diana Sachmpazidi}
\affiliation{University of Maryland, College Park, Maryland 20742, USA}
\affiliation{Western Michigan University, Kalamazoo, Michigan 49008, USA}

\author{Charles Henderson}
\affiliation{Western Michigan University, Kalamazoo, Michigan 49008, USA}

\author{Justyna P. Zwolak}
\email{jpzwolak@nist.gov}
\affiliation{National Institute of Standards and Technology, Gaithersburg, Maryland 20899, USA}

\date{\today}%
\begin{abstract}
Likert-style surveys are a widely used research instrument to assess respondents' preferences, beliefs, or experiences.
In this paper, we propose and demonstrate how network analysis (NA) can be employed to model and evaluate the interconnectedness of items in Likert-style surveys.
We explore the advantages of this approach by applying the methodology to the aspects of student experience scale dataset and compare the results to the principal component analysis.
We successfully create a meaningful network based on survey item response similarity and use modular analysis of the network to identify larger themes built from the connections of particular aspects.
The modular NA of the network of survey items identifies important themes that highlight differences in students' overall experiences.
Our network analysis for Likert-style surveys methodology is widely applicable and provides a new way to investigate phenomena assessed by Likert-style surveys.
\end{abstract}
\maketitle

\section{Introduction}\label{sec:introduction}
Surveys are important instruments for gathering data to answer questions about how many people experience and think about particular phenomena. 
For several decades, scholars, practitioners, and policymakers have been using national survey data to gather information about the state of our educational system and to take informed actions towards improving and evaluating the effectiveness of educational reforms~\cite{Desimore04-ARQ}.
In physics education research (PER), surveys have been developed to capture a wide range of student experiences, such as physics identity formation~\cite{Marshman17-SMC}, attitudes towards learning physics~\cite{Adams06-CLA}, and experiences with department support structures~\cite{Sachmpazidi21-DSS}.
These types of surveys are designed to be administered many times across different university and department settings to build conclusions with evidence from many contexts.
In analyzing large amounts of data, it is important to select the method of analysis that matches the research questions and provides useful insights into the larger phenomena under study.

Survey validity is typically assessed through confirmatory or exploratory analysis.
Principal component analysis (PCA) and exploratory factor analysis (EFA) are the most commonly used exploratory techniques.
In general, PCA aims to optimize the grouping of individual variables (here, the individual survey items), into a set of higher order {\it components}, which we call {\it survey thematic groups} or {\it themes}.
The goal of EFA is to identify survey themes (the so-called {\it hidden factors}), that explain or contribute to the observable variables (survey items).
These two approaches are similar but start with very different assumptions.
In PCA, the assumption is that there are larger themes that can be built from looking at the individual survey items.
In EFA, the assumption is that each of the individual survey items is the manifestation of a larger contributing thematic factor.

In the recent decades, network analysis (NA) has received increasing attention in the physical and social sciences~\cite{Borgatti09-NAS}.
In education research, NA has been used to understand the interactions between students both inside and outside of the classroom \cite{Thomas00-TTB, Forsman14-NAM, Forsman15-CSR, Grunspan14-UCS}.
Additionally, this approach has been used to understand the professional networks of instructors and the relation to their teaching practices \cite{Baker11-PBS, Siciliano16-QQT}.
In NA the assumption is that each individual variable (represented as a {\it node}), can be locally related to other variables (with the relationship represented as an {\it edge} connecting appropriate nodes).
The composition of nodes and edges makes up the network.
Similar to both PCA and EFA, clusters can be identified in the networks.
However, NA can provide additional insight about how exactly these clusters are formed and the importance of individual nodes within them.

In this paper, we introduce {\it network analysis for Likert-style surveys} (NALS), a method to evaluate and study Likert-style surveys through NA.
Previous research employing NA for exploratory analysis found that NA provides more detailed measures of interconnection between variables than EFA~\cite{Lee20-CDC}.
However, the variables used in that study as well as connections between them are predefined and established based on prior research.
NALS, on the other hand, relies on building a network from a single survey allowing for in-depth analysis of survey features.

We demonstrate the use of NA for Likert-style survey items by building a network from student responses to an established survey instrument~\cite{Sachmpazidi21-DSS}.
To validate our approach, we first demonstrate how survey items can be clustered using NA techniques.
Second, we highlight the differences and similarities of using NA and other traditional survey analysis methods (e.g., PCA).
We show that using NA to analyze the Likert-style survey reveals the role of individual survey items in relation to others as well as captures complexities in the building of larger themes, which is significantly different from the single-level grouping of PCA.
Finally, we demonstrate how survey validity and question redundancy can be assessed through NA tools.
To enable future use of NALS, we established a GitHub repository of the R source code, along with a manual and a toy example dataset, that can be easily adapted to other survey datasets \cite{na-likert}.

The manuscript is organized as follows: In Sec.~\ref{sec:background} we situate NALS in the context of other NA techniques used in PER and introduce the dataset we use.
In Sec.~\ref{sec:method} we introduce our methodology for creating a network of Likert-style survey items and the tools we use to analyze the network.
In Sec.~\ref{sec:results} we demonstrate our approach on a specific dataset in order to give an example of the types of findings and insights made possible by NALS.
Finally, in Sec.~\ref{sec:conclusion} we summarize the results and discuss the future work that could be accomplished through our proposed technique. 

\section{Background}\label{sec:background}
In this section, we review and compare the use of NA in PER.
We then introduce the dataset used to demonstrate the approach we propose as well as provide an overview of other types of analysis used in survey design.

\vspace{-5pt}
\subsection{Network analysis in PER}
In PER, NA has been used to investigate two primary domains.
The first focuses on the social and academic networks of physics students and how various network features relate to other factors of students or instruction \cite{Zwolak18-ECN, Williams19-LEP, Dou19-SNA, Traxler20-ALE}.
The students are the nodes and the edges are their interactions with their peers, either observed by researchers or self-reported by students.
Some studies seek to explain a particular phenomenon, such as  self-efficacy, persistence in introductory physics, or anxiety, based on their location in the network~\cite{Dou18-UDI, Dou16-BPM, Zwolak17-SNI, Vargas18-CSC}.
Other studies of in-classroom networks have been used to characterize learning environments, and used to evaluate the extent of interactions between students~\cite{Brewe12-PLC, Commeford21-ALE}.
Networks have recently been used in PER to explore different levels of interaction, including collaboration between lab groups and the effect of gender composition~\cite{Sundstrom22-LIG}.
This area of study focuses on the analysis and interpretation of student interactions and have shown how social network analysis can be used to uncover relations between social interaction and learning.

The second prominent area in which network analysis has emerged in PER is in studies related to concept inventories.
Concept inventories are standardized tests that aim to accurately assess the conceptual knowledge of students related to a specific physics topic.
The methodology, first developed by Brewe {\it et al.}, is known as module analysis for multiple choice responses (MAMCR)~\cite{Brewe16-MAM}.
While modular analysis is used in a wide variety of network science studies, the MAMCR enables the use of modular analysis for investigation of multiple choice responses.
The methodology aims to guide researchers in identifying groups of conceptually related ideas that are represented by student responses to multiple choice questions.
This approach was initially used to identify conceptual modules within the Force Concept Inventory (FCI), but has been expanded to other concept inventories, such as the Survey of Electricity and Magnetism and a quantum mechanics concept inventory~\cite{Wheatley21-CEM, Wells21-QMC}.
The method has been continually built upon and new interpretation has been introduced in order to identify alternative conceptual groupings and investigate the reported gendered differences in FCI performance~\cite{Wells19-MMA}.
These types of studies have been successful at classifying student responses into coherent conceptual structures through modular network analysis.

NALS is similar to MAMCR in that neither aim to analyze the social relations of individuals, but rather use their responses to investigate relationships among concepts or ideas.
However, MAMCR builds networks  on individual responses to items rather than on the survey items themselves.
NALS relies on building the connecting edges based on the similarity of responses to the survey items themselves.

In Likert-style surveys, participants respond with the level to which they agree or disagree with a particular statement representing a unique aspect of an experience or belief.
These statements constitute the survey items.
One could imagine that these survey items may be interrelated based on many attributes, such as how respondents collectively answer particular questions in similar ways.
In treating each survey item as a unique unit of analysis, we aim to investigate the manner in which survey items are related to each other from the perspective of the survey respondents.
NA is unique when compared to PCA and EFA in that it relies on the local connections of individual nodes (variables) to capture features of the overall phenomenon. 
In using NA to study Likert-style surveys, we see an opportunity to better map the connections that certain survey items have to each other, identify those survey items that drive connections between other items, and characterize the resulting networks by the types of clusters that form. 
This approach is uniquely based on how respondents themselves report connections between survey items, rather than identifying collective trends first as in PCA, and includes a wide set of tools to analyze those connections.

\vspace{-5pt}
\subsection{The ASES instrument}\label{ssec:ases}
To provide an example of the types of questions and findings that can be asked and uncovered through these methods, the aspects of student experience scale (ASES) dataset is used \cite{Sachmpazidi21-DSS}.
ASES is an instrument intended to assess physics graduate student experience of departmental support structures.
The initial items were developed based on prior literature and the American Physical Society Bridge Program recommendations for creating a supportive and complete educational experience to Bridge students~\cite{apsbp}.
The utility of the ASES was assessed through responses of 397 students from 19 physics graduate programs.
PCA revealed four components: mentoring and research experience ($E$), professional development ($D$), social and academic integration ($I$), and financial support ($S$).

A later study used ASES data to examine a graduate retention model using structural equation modeling~\cite{Sachmpazidi21-PhD}.
The authors found the critical role of social and academic integration as well as mentoring and research experience in predicting academic self-efficacy and intention to persist.
These results suggest the utility of ASES for understanding graduate student experience and retention.

\begin{figure*}[ht]
    \includegraphics{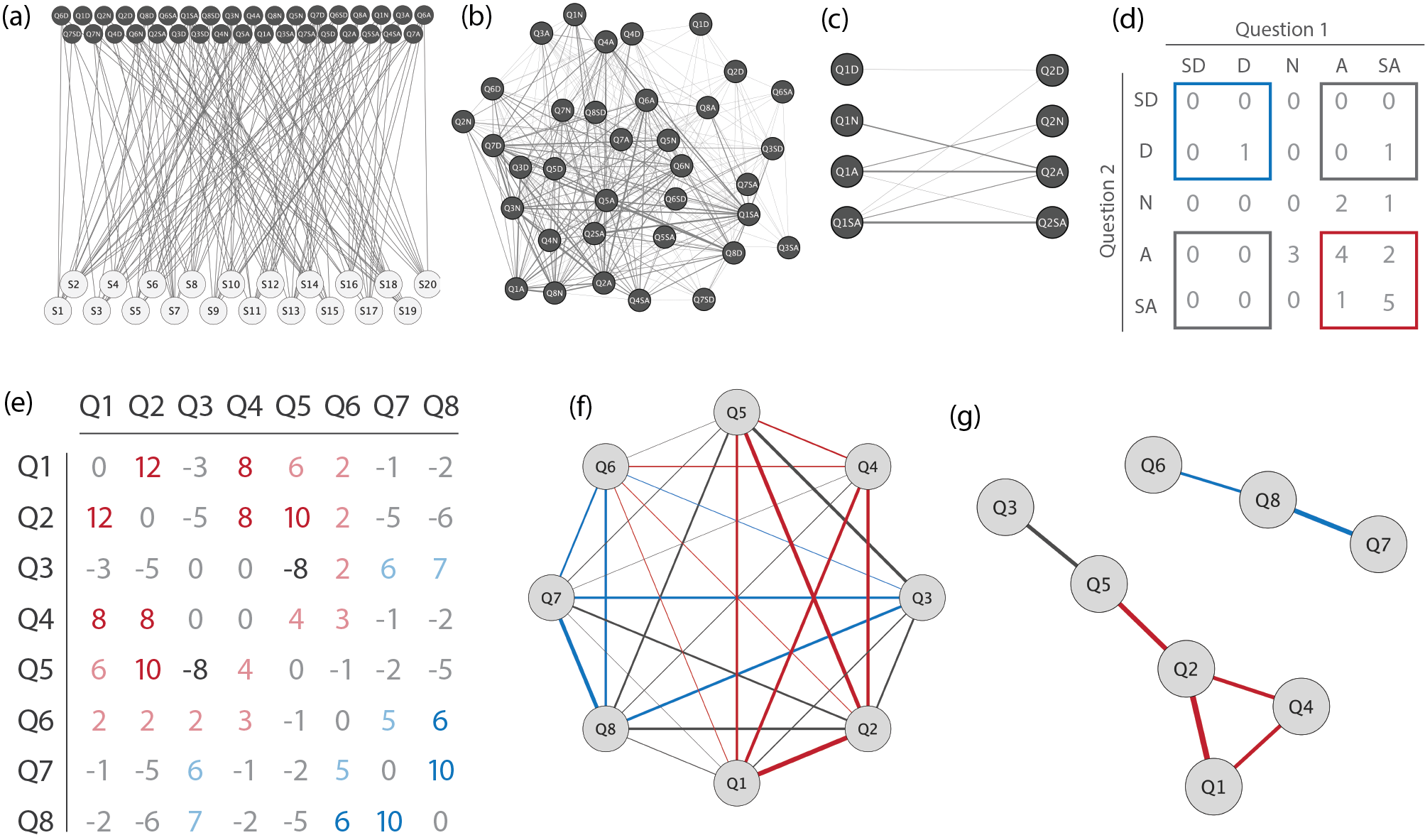}
    \caption{Visualization of generating a network based on Likert-style survey responses. 
    (a) Bipartite network of students (light gray) and response selections (dark gray).
    (b) Projection of bipartite network onto the response selections, weighted by number of students connecting. 
    (c) Network of response selections for just Q1 and Q2.
    (d) Adjacency matrix representation of Q1 and Q2 response network with the negative temperature relationships boxed in blue, the positive temperature relationships boxed in red, and the inverse relationships boxed in gray.
    The same color scheme is used in the remaining panels.
    (e) Adjacency matrix of the survey item network after collapsing the response network with significant edge weights in bold.
    (f) A full survey item network with all edges included. 
    (g) A backbone survey item network with only significant edges.}
    \label{fig:model-net}
\end{figure*}

\vspace{-5pt}
\subsection{Principal component analysis}\label{sec:pca}
In most studies, researchers use multiple variables to explain a phenomenon.
It is often likely that some of the used variables end up conveying similar information, therefore these variables could instead be represented by fewer variables that are necessary to understand the underlying concept.
In these cases, researchers apply methods suitable for dimensionality reduction \cite{Demsar13-PCA}.
Principal component analysis (PCA) is the most widely used method for dimensionality reduction \cite{Candes11-PCA}. 
 
PCA is a method by which researchers transform combinations of the original variables in a dataset (e.g., student Likert responses) such that the new combinations capture the maximum possible variance in the dataset while eliminating correlations.
To achieve this outcome, first a covariance matrix of the original (mean-centered) data is created.
The eigenvectors and eigenvalues of this matrix are then used to transform the original data into a new set of variables.
The eigenvectors define the directions of the principal components and the eigenvalues define the magnitude of each eigenvector.
Finally, the principal components with the highest eigenvalues are retained as principal component since they account for the most variance in the data.

Additionally important to PCA is the interpretation and eventual naming of the principal components identified.
In this step, the disciplinary expertise and experience of the researchers is vitally important.
The researchers bring their knowledge to interpreting what common theme brings the survey items together.
This introduces a human element into the research process and can be executed proficiently or poorly, contributing to the overall validity of the work.

\section{Methodology}\label{sec:method}
In this section, we describe the processes of creating and analyzing the backbone survey item network.
The network creation is given special attention as the approach is innovative and unique to Likert-style survey data. 
The analysis described focuses on modularity analysis and node-centric centrality measures.

\vspace{-5pt}
\subsection{Creation of the network}\label{ssec:creation}
One way to build a network from responses to Likert-style surveys is by quantifying the similarity in how respondents answer questions.
We build the connections between survey items based on the level of agreement between students answering any two questions on the survey.
The process of building the backbone  survey item network is depicted in Fig.~\ref{fig:model-net}.
Note that the underlying assumption of NALS is that all survey items are coded the same direction (i.e., selecting agree as desirable).
Thus, data from reverse coded questions should be reordered prior to analysis.

The results from the survey are first converted into a {\it bipartite network} structure.
Bipartite networks are defined by two sets of nodes, one indicating respondents and the other being all possible responses to survey questions, that have connections only with nodes of the opposite set.
In these data, the students who participated in the survey make up group A, and the individual Likert scale selections make up group B, with edges indicating each students' selections.
Figure~\ref{fig:model-net}(a) depicts a toy example, in which we have a pool of 20 students responding to eight survey questions with five possible selections (strongly disagree, disagree, neither agree nor disagree, agree, and strongly agree), leading to 40 possible response selections with a single student connected to eight of these.

The next step in generating the network is to {\it project} the bipartite network onto the response choices [as shown in Fig.~\ref{fig:model-net}(b)].
Figure~\ref{fig:model-net}(c) takes a closer look at the connections between responses to Q1 and Q2 only.
In our toy example, a particular student selects strongly agree to Q1, and agree to Q2, which results in an edge in the projected network between Q1SA and Q2A.
That same student also selects agree with survey item Q3, resulting in Q3A connecting to both Q1SA and Q2A in the projected network, and so on.
A second student also selects strongly agree to Q1, agree to Q2, but selects neither agree nor disagree to Q3.
Thus, the weight of the edge between Q1SA and Q2A becomes 2.
However, since the second student selects neither agree nor disagree to Q3, the weight of the other edges remain 1.
Thus, in the projected network of responses, the weights of the edges represent the number of students who selected both of the connected responses.
We can see the result of this step in our model in Fig.~\ref{fig:model-net}(b), in which the response selections are projected in a network, with thicker edges relating to larger weights.

To obtain a network of survey items, rather than just the Likert scale responses, we start with the {\it adjacency matrix}.
The adjacency matrix of a network is an $N\times N$ matrix, where $N$ is the total number of nodes in the represented network.
Each element of this matrix holds information about the connection between the nodes represented by the row and column corresponding to that element.
In the adjacency matrix for the Likert scale selection projection, each element represents the number of respondents that selected the two responses associated with that particular row and column (e.g., if two students selected strongly agree to Q1 and selected agree to Q1, the element in column Q1SA and row Q1A would be equal to 2).

In the first step in calculating a single value for the weight between survey items, the adjacency matrix is split into submatrices associated with each possible pairing of survey items.
For our model, this results in 56 $ 5\times 5$ submatrices, since each of the eight survey items could be connected to the other seven items.
Since the adjacency matrix is symmetric by design, only 28 submatrices are unique.
The rows and columns of each matrix represent the weight of the edge between two unique Likert scale selections associated with two different survey items.
Elements along the diagonal show the absolute similar selections between the two survey items, e.g., Q1A and Q2A.
An example of a submatrix built from the network depicted in  Fig.~\ref{fig:model-net}(c) is shown in Fig.~\ref{fig:model-net}(d).

The goal of the final step is to collapse each submatrix into a single value that represents the edge weight between the two survey items.
In our approach, we treat both forms of disagreement (i.e., disagree and strongly disagree) as well as both forms of agreement (i.e., strongly agree and agree) as indication of similar attitudes.
First, all of the matrix elements that represent similar responses are summed [see the diagonal red and blue boxes in Fig.~\ref{fig:model-net}(d) which sum up to 13].
Then, all of the matrix elements that represent dissimilar responses are summed [see the off-diagonal gray boxes in Fig.~\ref{fig:model-net}(d) which sum up to 1].
The final weight is obtained by subtracting the off-diagonal sum from the diagonal sum (resulting in 12 for the example matrix).
The elements along the neutral row and column are ignored in this calculation as they indicate lack of attitude to a given question.
The result is a single value that represents the similarity score between two particular survey items.
A positive value indicates that the two items are answered with similar responses.
A negative value indicates that the two items are answered with dissimilar responses.
If the result is 0, then no edge is created between these two items, indicating no distinct relationship between response selections.
This is repeated for all pairs of survey items.
An example of the resulting adjacency matrix for our model is shown in Fig.~\ref{fig:model-net}(e).

We use the edge weight in the full survey item network to represent the similarity of the Likert scale selections.
However, the edge weight alone does not indicate the level of agreement, i.e., whether two survey items are connected through mutual selections of disagree or agree.
Therefore, we introduce an additional edge attribute, dubbed {\it temperature}.
Temperature is an attempt to capture the difference between the amount of agrees and the amount of disagrees for each pair of survey items.
That is, temperature is calculated by subtracting the sum of all mutual disagrees from the sum of all mutual agrees.
Temperature is a continuous measure that ranges from $-n$ to $+n$, where $n$ is the total number of respondents in the survey.
Thus, two nodes that are often both disagreed with will share an edge with a negative temperature value (blue), while two nodes that are often both agreed with will share an edge with a positive temperature (red).
Through the temperature value of the edge, information about the level of agreement between responses that connect two survey items is retained.
Figure~\ref{fig:model-net}(e) shows the resulting color corresponding to temperature for each connection in our toy example.

The resulting survey item network is extremely dense, with almost every node connected to every other node, as seen in Fig.~\ref{fig:model-net}(f).
In order to preform analysis of the network, we must identify the most important connections between each node.
In NA, this involves identifying the so-called {\it backbone} of a network.
The backbone is the residual network after eliminating non-significant edges.

A number of algorithms have been developed to extract the backbone of dense networks.
In previous NA done in PER, the locally adaptive network sparsification (LANS) algorithm was employed to find the network backbone~\cite{Brewe16-MAM}.
In this approach, the significant edges for each node are identified and kept in the network.
Significance is defined through an empirical cumulative distribution function applied to the edge weights associated with each node~\cite{Foti11-NSN}.
The level of significance, $\alpha$, is used to determine the percentage of edges kept during sparsification ($1-\alpha$).
This is set based on the size of the network under consideration, typically $\alpha=0.05$ for smaller networks and $\alpha=0.01$ for larger networks.
This approach keeps all edges that are significant for at least one node of that pair.

We implement the LANS algorithm to extract the weighted backbone of the survey item similarity network in order to identify the connections that were prevalent for the highest number of respondents.
The absolute value of the edge weight used in the LANS algorithm allows us to account for the weight of edges both due to the positive value (more similarities than differences) as well as the negative value (more differences than similarities).
The significant edges in our model are shown in bold in the adjacency matrix in Fig.~\ref{fig:model-net}(e).
The final network is visualized in Fig.~\ref{fig:model-net}(g).
This final backbone survey item network is made up of two components.
The larger component contains one edge that represents an inverse relationship and four edges indicating similar relationships built through agree responses.
The smaller component is made up of two edges representing similar relationships built through disagree responses.

Through applying the steps of the methodology described above, we are able to create a network based on the significant similar responses to Likert-style survey items.
The backbone survey item network is the outcome of the construction phase which is then used in the analysis.
The source code for generating the network following the methodology described in this section is available on a GitHub repository~\cite{na-likert}.
In the next section, we describe our approach to analyzing the backbone survey item network.

\vspace{-5pt}
\subsection{Analysis of the network}
One may be interested in both how nodes group together within the survey item network as well as which individual nodes have the highest importance within this network.
To understand how nodes group together, a partitioning process is needed to detect the clusters of survey items that exist in the created network.
We perform this analysis using the backbone survey item network resulting from the above construction process.
Such clusters are defined as having significantly more internal edges compared to external connections.
The measure of how well the network can be partitioned is called {\it modularity}.
The partitioning of our network optimizes the modularity in the identification of survey item clusters.
In our analysis, we have made use of the weighted modularity as defined by Newman~\cite{Newman04-AWN}.

The modularity is given by
\vspace{-5pt}
\begin{equation}
    Q = \frac{1}{2m}\sum_{i,j} \bigg[w_{ij} - \frac{C_{S,i} C_{S,j}}{2m}\bigg]\delta(c_i,c_j),
    \vspace{-3pt}
\end{equation}
where $w_{ij}$ represents the weight of an edge between nodes $i$ and $j$, $C_{S,i}$ ($C_{S,j}$) represents the strength of node $i$ ($j$), $c_{i}$ ($c_{j}$) represents the cluster to which node $i$ ($j$) belongs, and $m = \frac{1}{2}\sum_{i,j}w_{ij}$.
The delta function, $\delta(c_i,c_j)$, equals $1$ when $c_i = c_j$ and $0$ otherwise.
The modularity ranges from $-1$ to $1$, with a positive value indicating a successful partition and a negative value indicating a weak partition.

To partition the backbone survey item network, we chose to implement the hierarchical clustering algorithm proposed by Clauset, Newman, and Moore, colloquially dubbed the {\it fast-greedy algorithm}~\cite{Clauset04-FCS}.
This algorithm follows a modularity optimizing process and produces a hierarchical ordering that identifies high level clusters along with the hierarchical relationships between individual nodes within each cluster.
This approach not only ensures finding the most optimal partition of the network but also provides insights into how each node individually fits within the clusters.

To compare the network partition introduced above with more traditional processes for grouping survey items (such as PCA) we use the {\it F-measure}~\cite{Ghawi22-CMS}.
The F-measure is made up of two quantities: {\it precision} and {\it recall}.
Precision quantifies the extent to which a cluster from partition $A$ contains items from only a single cluster in the partition $B$.
Recall captures the fraction of items that come from the cluster in partition $A$ out of the total number of items in the cluster of partition $B$.
The F-measure ranges from $0$, indicating no overlap in partitions, to $1$, indicating a perfect matching.
Traditional cluster comparison is asymmetric, treating one partition as the base partition and the other as the variant.
As proposed by Ghawi and Pfeffer~\cite{Ghawi22-CMS}, we use two-way comparison and find the combined F-measure as the harmonic mean of each component.

The clustering techniques can be also used to evaluate validity of surveys and identify redundant questions.
This type of analysis is important for survey designers in order to ensure that the instrument is measuring the desired underlying concepts.
In the language of modular analysis, the validity can be captured through the coherence of clusters.
What this means in practice is that through the interpretation of the clusters, researchers can evaluate if the intended themes of their surveys are coherently captured by the clusters and how the survey items are sorted.
Additionally, due to the network model, NALS can help researchers understand the complexities within the identified themes themselves and the differences between themes in terms of how they are built from the survey items.
For example, one theme may be more loosely held together through weaker connections than another theme.

Using these same techniques, researchers can remove periphery items from the survey based on their location in the hierarchical ordering: the higher up in a cluster a particular survey item is, the less central that item is to the larger theme represented by that cluster.
This feature is important for researchers as one may be interested in what survey items are most central to the intended themes.
NALS provides a tangible way to identify which survey items are most and which are least central.
Researchers may use the latter information to eliminate questions that are not central to an identified theme.
Additionally, they may seek to edit or add survey items to increase the capacity to measure a particular theme.

A complimentary way to quantify the relative importance that individual nodes hold within a network is through centrality measures.
We aim to consider both traditional centrality measures of \textit{degree} and \textit{strength}.
The degree of a node is the total number of edges that are attached to that node, while the strength of a node is calculated as the sum of all of the weights of those edges.
A node may have a high degree while maintaining a low strength (many weak connections); alternatively, a node may have a low degree but a high strength (few strong connections).
To capture both of these features, we use the weighted degree as defined by Opsahl {\it et al.}~\cite{Opsahl10-NCW}.
The weighted degree has been chosen in order to account for both the number of connections between each survey item and also the weight of those connections.

The weighted degree is defined as follows:
\vspace{-3pt}
\begin{equation}
    C_{D,i}^{\,\beta} = (C_{D,i})^{1-\beta}(C_{S,i})^{\,\beta},
\end{equation}
where $C_{D,i}$ is the traditional degree of node $i$, $C_{S,i}$ is the strength of node $i$, and $C_{D,i}^{\,\beta}$ is the weighted degree of node $i$.
The value of $\beta$ acts as a tuning parameter for how degree and strength are valued in the calculation of weighted degree.
If $\beta = 0$, then only degree is considered while for $\beta = 1$ only strength is considered.
In between $0$ and $1$, the degree and strength are both included in the calculation of the weighted degree.
Since we are interested in both of these values, we have chosen $\beta = 0.5$ to give equal importance to both degree and strength.

The weighted degree can also be used as a heuristic for survey validity and item removal.
This centrality measure captures the extent that respondents view a particular item as similar to others in the survey.
Thus, a low weighted degree indicates an item that is responded to in unique ways.
This can act as a check for survey designers to understand whether respondents are reacting to items in expected or unexpected ways.
Depending on the research objective, degree can be used to identify items that are candidates for removal by being either unlike the other items (i.e., only weakly related to the underlying concept of the survey) or being too similar to the other items (i.e., not providing additional information to what is captured by other survey items).

\vspace{-5pt}
\subsection{Visualization and statistical analysis}\label{ssec:viz-stat}
All analyses and dendrogram visualization presented in this work are carried out using the {\it igraph} package in {\it R} \cite{igraph, R}.
The network visualization is created using the open source software {\it cytoscape} \cite{cytoscape}.
Employing the LANS algorithm, we use a level of significance $\alpha = 0.05$.

\section{Results}\label{sec:results}
\begin{figure*}[!ht]
    \includegraphics{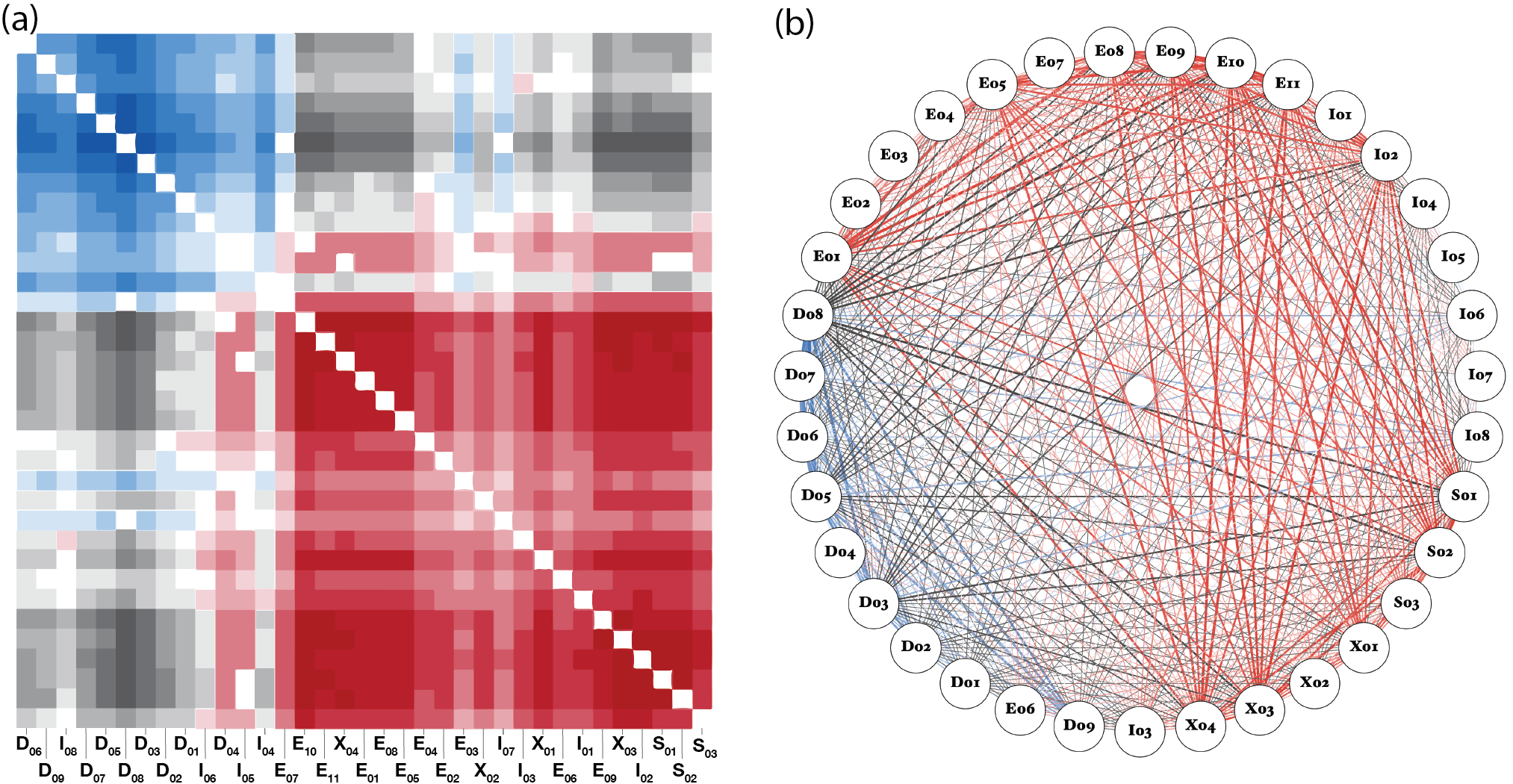}
    \caption{Representations of the full network of ASES items. (a) Adjacency matrix of ASES items represented by a heat map. (b) Network of survey items arranged in a circle with edge weights corresponding to the absolute value of the similarity. Both the adjacency matrix representation and the network representation use two scales: one for items that share similar responses that ranges from red (mutual agreement) to blue (mutual disagreement), the second for items that share dissimilar responses ranging from black (most dissimilar) to white (no correlation in responses).}
    \label{fig:ASES-full}
\end{figure*}

The validation of the NA methods for Likert-style surveys put forward in Sec.~\ref{sec:introduction} can be restated as three high-level research questions:
\begin{enumerate}[nosep]
    \item How does modular analysis group survey items?
    \item How does modular analysis of networks compare to PCA and the components identified?
    \item Can we use NA to validate the survey design and identify redundant questions?
\end{enumerate}

To answer the first research question, we employ the proposed methods to create a network representation of the ASES dataset described in Sec.~\ref{ssec:ases} and identify the underlying groups of experiences.
The second question motivates a comparison to results from PCA, which we evaluate quantitatively.
A qualitative interpretation of the differences is presented in Sec.~\ref{sec:inter}.
Finally, to address the third question, we propose and model heuristics for survey validity and item removal.

\subsection{How does modular analysis group survey items?}
\begin{figure*}[!ht]
    \includegraphics{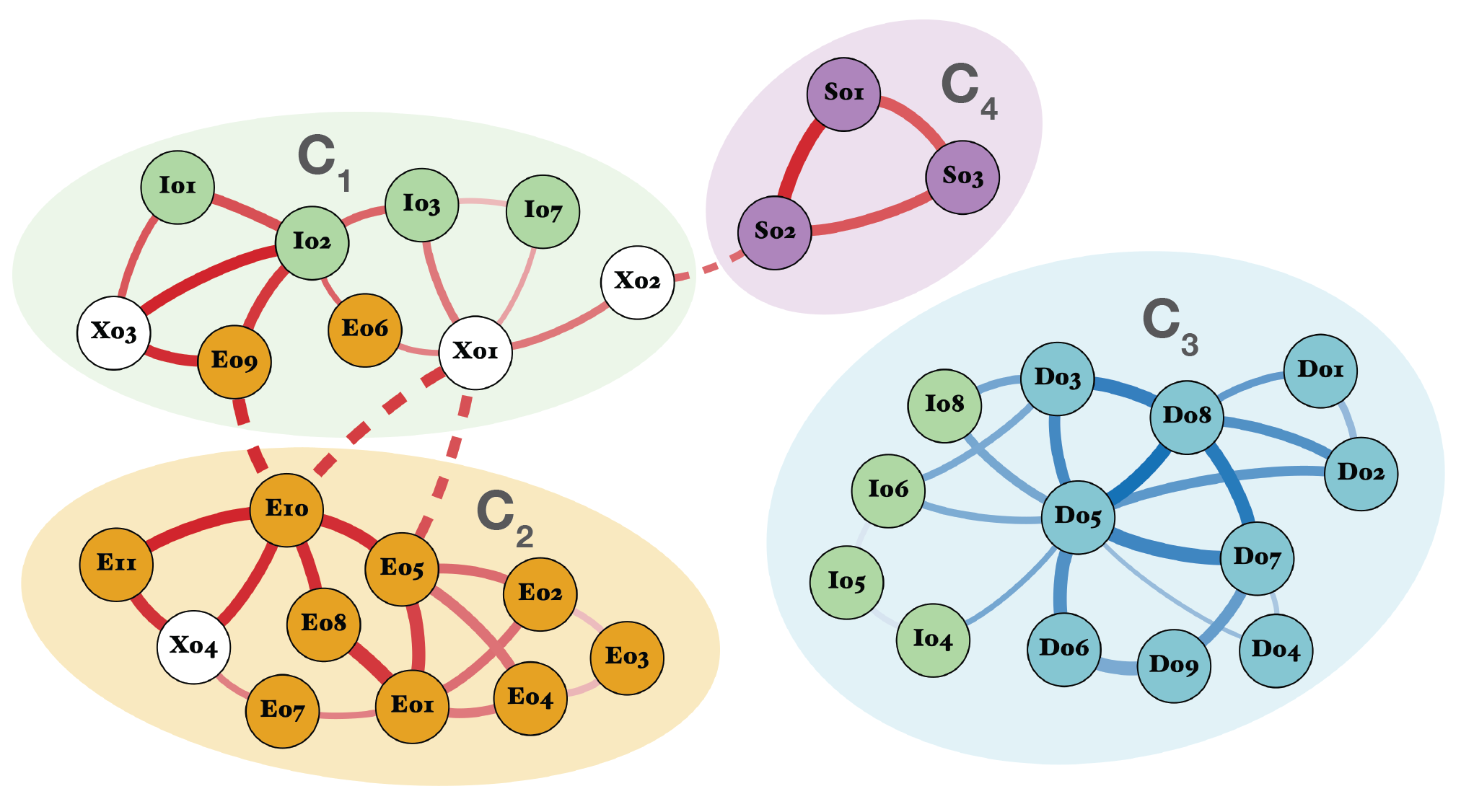}
    \caption{The ASES similarity network. The node colors indicate the principal component that the survey item was loaded into; mentoring and research experience, $E$ (orange), professional development, $D$ (teal), social and academic integration, $I$ (green), financial support, $S$ (purple), or not loaded (white). The shaded regions represent the clusters identified by the fast-greedy clustering algorithm; $C_1$ (green), $C_2$ (orange), $C_3$ (teal), or $C_4$ (purple). The edge thickness represents the amount of similarity. The edge color indicates the edge temperature, i.e., whether the edge is created via mutual agree (red) or disagree (blue) answers. The dashed edges indicate connections that span across clusters.}
    \label{fig:ASES-net}
\end{figure*}

The ASES is made up of 35 questions, each with five possible response options: strongly disagree, disagree, neither agree nor disagree, agree, and strongly agree.
Thus, there are a total of 175 unique responses possible.
Each of the 397 student respondents is connected to 35 of these 175 options in the bipartite network.
This bipartite network is then projected onto the possible responses, as described in Sec.~\ref{ssec:creation}, resulting in a maximum possible edge weight of 397. 

The projected network of options is then collapsed to create a network of survey items.
The size of the full adjacency matrix used to perform this step is $175\times 175$, with a total of 595 submatrices used to calculate similarity and temperature between items.
The resulting collapsed adjacency matrix ($35\times 35$) is represented as a heat map in Fig.~\ref{fig:ASES-full}(a).
Similar to the toy example, the ASES network is very dense, with almost every survey item connected to all others.
This dense network, visualized in Fig.~\ref{fig:ASES-full}(b), contains connections built on mutual agreement (positive similarity and positive temperature; red edges), mutual disagreement (positive similarity and negative temperature; blue edges), and inverse response selections (negative similarity; gray edges).
We use Fig.~\ref{fig:ASES-full}(b) to represent the high density of the full survey item network and to motivate the need of sparsifying it prior to the modular analysis. 
Figure~\ref{fig:ASES-net} shows the sparsified backbone survey items network.
While there are indications of different groups of items based on types of connections (positive and negative temperature), the dense network makes it difficult to identify specific clusters.

To investigate the network through a cluster analysis and other NA techniques, we first identify the significant connections within the dense network through the LANS algorithm.
The resulting backbone survey item network includes 35 nodes (one for each ASES item) and 55 significant edges, out of which 34 ($62~\%$) have a positive temperature and 21 ($38~\%$) have a negative temperature (Fig.~\ref{fig:ASES-net}).
Moreover, the network is split into two subnetworks that correspond to the temperature, with the positive temperature subnetwork consisting of 22 survey items (component A) and the negative one consisting of 13 items (component B).

\begin{figure*}[t]
    \includegraphics{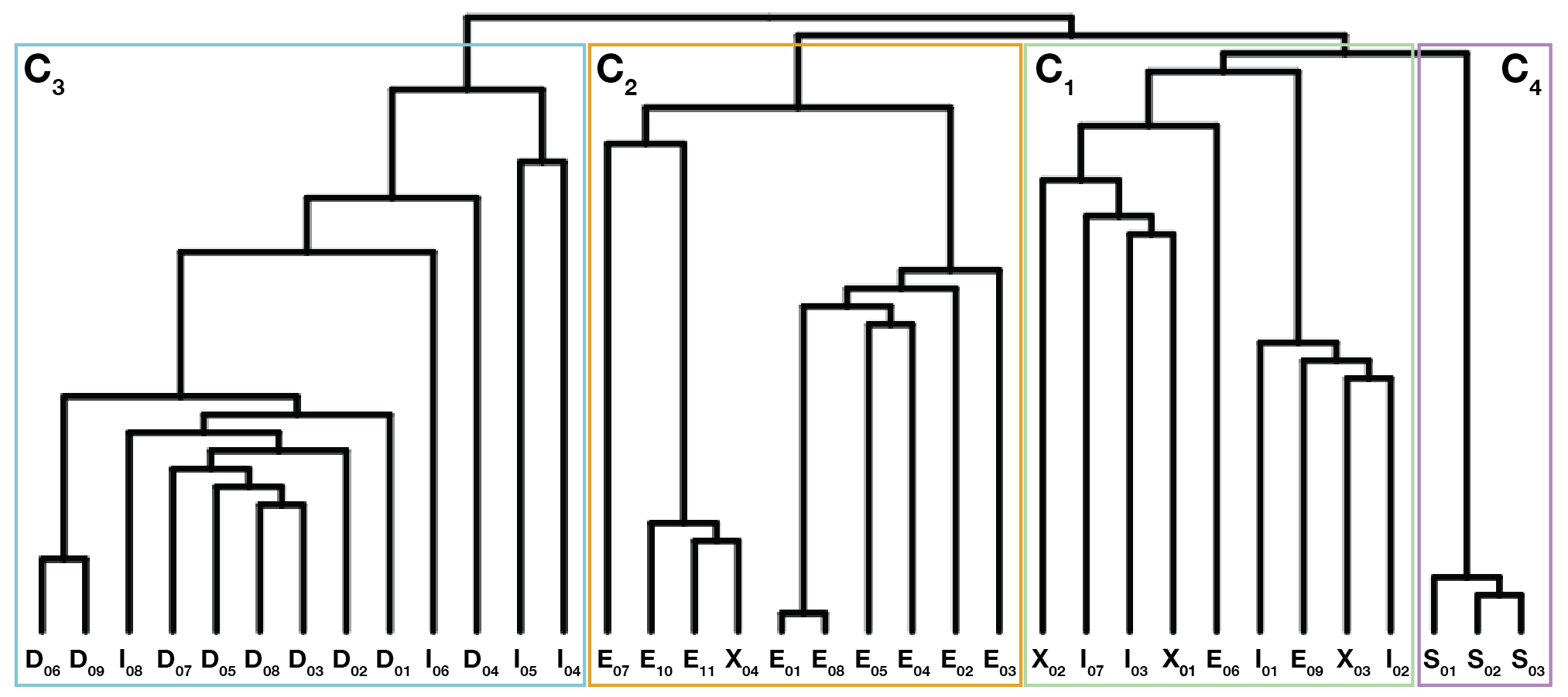}
    \caption{The hierarchical structure of the partition provided by the fast-greedy cluster detection algorithm, represented as a dendrogram. 
    Each node name is listed along the bottom. 
    Each of the four clusters is identified by colored boxes around the respective branches; $C_1$ (orange), $C_2$ (green), $C_3$ (teal), or $C_4$ (purple).}
    \label{fig:ASES-dendro}
\end{figure*}

\vspace{-5pt}
\subsection{How does NA compare to PCA?}
The next step in the analysis involves identifying the partition of this network through the fast-greedy clustering algorithm.
In the ASES backbone survey item network, four clusters were detected (Fig. \ref{fig:ASES-net}).
Component $B$ is partitioned into a single cluster, while component $A$ is separated into three clusters.
The modularity is $0.62$, indicating a meaningful split of the network. 
The network clustering is depicted in Fig.~\ref{fig:ASES-net}.

The first cluster, $C_1$, is made up of nine survey items.
This cluster is a mixture of two principal components, with four items belonging to PCA component $I$ and two to PCA component $E$.
In addition, three items not loaded into a component by PCA are also grouped in this cluster.
The second cluster, $C_2$, is made up of ten survey items.
$C_2$ is very similar to the PCA component $E$, with nine items partitioned from $E$ into this cluster.
The tenth item in $C_2$ is one of the items not loaded into a PCA component.
The third cluster, $C_3$, encompasses all of PCA component $D$ along with four items from PCA component $I$.
Finally, the fourth cluster, $C_4$, groups the exact same items as PCA component $S$.
It is the smallest cluster.

As evident from Fig.~\ref{fig:ASES-net}, while there are some similarities in how NA and PCA group the survey items, there are some important differences.
In particular, the calculated F-measure between the partitions from PCA and NA is 0.75, indicating similarity between the principal components and the clusters.
However, there are some interesting differences, which we discuss in Sec.~\ref{sec:inter}.

Another way to understand the partitions generated through the fast-greedy algorithm is through a dendrogram, as seen in Fig.~\ref{fig:ASES-dendro}.
In this representation, each node shown at the bottom of the figure is connected to other nodes through hierarchical branches.
The lower a node is connected through a branch, the more essential that node is to the cluster.
Using the dendrogram, we can see that both $C_1$ and $C_2$ contain two subgroups of about equal number of nodes, whereas $C_3$ has a core group of nodes with few nodes being added higher on in the hierarchy.
This is somewhat analogous to the loading factors provided by PCA.

\subsection{Can we use NA to validate the survey design?}
Building on the hierarchical clustering analysis we suggest heuristics that can be used as a guide in future survey development.
In using the resulting hierarchy, as seen in Fig.~\ref{fig:ASES-dendro}, one can better understand how the larger themes are built from the ordering of survey items through identifying core nodes to each cluster or observing important subclusters.
This, in turn, might reveal nodes---e.g., those added into clusters higher up in the structure---that can be pruned without affecting cohesion of the themes.
This approach is best suited for situations in which there is a core group of nodes in a cluster, rather than subclusters.
For example, in the ASES network, we may identify a number of possible items for elimination from $C_3$ that are added to the cluster higher up in the hierarchy and thus less central to the formation of this cluster (e.g., $I_{04}$ and $I_{05}$).

\begin{table}[b]
\renewcommand{\arraystretch}{1.05}
\renewcommand{\tabcolsep}{6pt}
\caption{\label{tab:tab-deg}%
Weighted degree, $C_{D,i}^\beta$, of each survey item, $i$, grouped by clusters. Here, $\beta=0.5$.}
\begin{ruledtabular}
\begin{tabular}{cc | cc | cc | cc}
\multicolumn{2}{c|}{Cluster $C_1$} & \multicolumn{2}{c|}{Cluster $C_2$} & \multicolumn{2}{c|}{Cluster $C_3$} & \multicolumn{2}{c}{Cluster $C_4$}                \\
$i$    & $C_{D,i}^\beta$  & $i$    &  $C_{D,i}^\beta$    & $i$       &  $C_{D,i}^\beta$          &  $i$ & $C_{D,i}^\beta$ \\ \hline
$I_{01}$     & 25.8          & $E_{01}$     & 68.8          & $I_{04}$        & 21.2       & $S_{01}$  & 29.7       \\
$I_{02}$     & 65.3          & $E_{02}$     & 37.5          & $I_{05}$        & 21.2       & $S_{02}$  & 41.2       \\
$I_{03}$     & 35.1          & $E_{03}$     & 22.9          & $I_{06}$        & 33.9       & $S_{03}$ & 27.9       \\
$E_{06}$     & 21.3          & $E_{04}$     & 38.5          & $D_{01}$        & 23.5       &                          &            \\
$X_{01}$     & 73.4          & $E_{05}$     & 70.6          & $D_{02}$        & 38.7       &                          &            \\
$X_{02}$     & 23.6          & $X_{04}$     & 40.7          & $D_{03}$        & 53.7       &                          &            \\
$X_{03}$     & 40.4          & $E_{07}$     & 22.4          & $I_{08}$        & 24.9       &                          &            \\
$E_{09}$     & 42.8          & $E_{08}$     & 30.7          & $D_{04}$        & 19.5       &                          &            \\
$I_{07}$     & 20.5          & $E_{10}$     & 87.6          & $D_{05}$        & 120.0      &                          &            \\
        &               & $E_{11}$     & 29.6          & $D_{06}$        & 30.7       &                          &            \\
        &               &         &               & $D_{07}$        & 56.1       &                          &            \\
        &               &         &               & $D_{08}$        & 72.4       &                          &            \\
        &               &         &               & $D_{09}$        & 29.6       &                          &     \\      
\end{tabular}
\end{ruledtabular}
\end{table}

This approach also can be used to identify themes that are more complex within the survey.
For example, in the ASES network, we see that both $C_1$ and $C_2$ have two clear subclusters, which indicates complex themes that bring more than one type of experience together.

Alternatively, we may consider the local centrality measure of the weighted degree that was calculated from the resulting backbone survey item network, as reported in Table \ref{tab:tab-deg}.
A larger $C_{D,i}^\beta$ value indicates that the particular survey question is strongly similar to many other items, and a smaller $C_{D,i}^\beta$ value means that the particular question is less similar to other items.
For example, in the cluster $C_1$, $X_{01}$ ($C_{D,X_{01}}^\beta$ = 73.4) is essential for the construction of this cluster; however $I_{07}$ ($C_{D,I_{07}}^\beta$ = 20.5) could be considered redundant and unnecessary in the cluster structure.
On the other hand, a low centrality indicates less similarity to other survey questions, meaning that the particular item is responded to in unique ways and captures different experiences.
Depending on the purpose of pruning, one might want to keep such items and remove some of the high centrality nodes instead.
Other heuristics could be used for the identification and removal of survey items.
The choice of methodology for survey pruning needs to be dependent on the survey purpose and should be guided by the research objective.

\section{Interpretation and discussion}\label{sec:inter}
NALS provides us with new insights that are not accessible through PCA.
While PCA results in a high-level partition of the survey items, this method does not consider the hierarchical dependencies between items within each partition.
NA, on the other hand, reveals the nuanced hierarchical relationships of items that are used to build each cluster from the bottom up.
That is, NA captures the multilevel complexities of the larger themes in addition to the high-level partitioning.

Additionally, NA allows us to quantify the relative dominance of individual items both within and between clusters.
These features are important for understanding how single survey items are answered in unique or highly similar ways.
The individual characteristics of items is not accounted for in PCA.

To interpret the findings from the network perspective, the actual survey that the network represents must be considered.
For the ASES instrument, the structure of two separate subnetworks indicates that there is an important division in the student experience of supportive departments.
The aspects in the subnetwork created through mutual agreement (positive temperature) can be seen as reinforcing an experience of a supportive department.
Alternatively, the aspects that are a part of the subnetwork made up of mutual disagreement (negative temperature) are indicative of support structures that many students do not experience.
While the negative temperature subnetwork makes up a single cluster, the subnetwork of positive temperature consists of three clusters.

\vspace{-10pt}
\subsection{Sensemaking of the NA clustering}
We have compared the clusters with the PCA components through the F-measure, which quantifies the level of similarity between the two partitionings, and found that there is a large amount of overlap between the two methods.
However, there are important differences that can be explored qualitatively.
As mentioned in Sec.~\ref{sec:pca}, the disciplinary expertise of researchers is important to interpret these partitionings.
Our research team brings our knowledge and understanding of both the ASES instrument and physics graduate programs in order to make sense of our results.

One of the most striking differences to come out of the two approaches is that the PCA component $I$, social and academic integration, is divided in the NA approach, as seen in Fig.~\ref{fig:ASES-net}.
In particular, half of the survey items that belong to this component are connected with items from the PCA component $D$, professional development---$I_{04}$, $I_{05}$, $I_{06}$, and $I_{08}$.
These four survey items are more aligned with the type of academic coursework support that students may experience, as reflected in the $C_3$ cluster in Fig.~\ref{fig:net-words} (additionally, see Appendix~\ref{app:clustering}).
Being grouped into the negative temperature cluster, $C_3$, means that these support structures, along with those related to professional development, are not often experienced by respondents.
Thus, the aspects in $C_3$, which we call {\it professional and academic development}, form a group of support constructs that are lacking from many students' experiences (Fig.~\ref{fig:net-words}).

The other four aspects in the PCA component $I$ are connected with two items from the PCA component $E$, mentoring and research experience, as well as three survey items that were not loaded into any component through PCA.
The four survey items that come from component $I$ are closely related to social and peer support.
The two survey items from component $E$ are related to the social support and flexibility that one may experience as part of graduate research.
The three factors that were previously not loaded are focused on the exploratory aspects of a graduate degree.
These items, forming cluster $C_1$, indicate supports that are mutually reinforcing and center the student experience around social and exploratory aspects within a department.
We call $C_1$ {\it social and scholarly exploration support}, as seen in Fig.~\ref{fig:net-words}.

Network clusters $C_2$ and $C_4$ are very similar to the two PCA components $E$ and $S$, mentoring and research experience and financial support, thus each cluster is named based on the corresponding PCA component (Fig.~\ref{fig:net-words}).
While the differences in the two approaches is exemplified in $C_1$ and $C_3$, $C_2$ and $C_4$ make it clear that the par-

\onecolumngrid

\begin{figure*}[!hb]
    \includegraphics[width=0.98\linewidth]{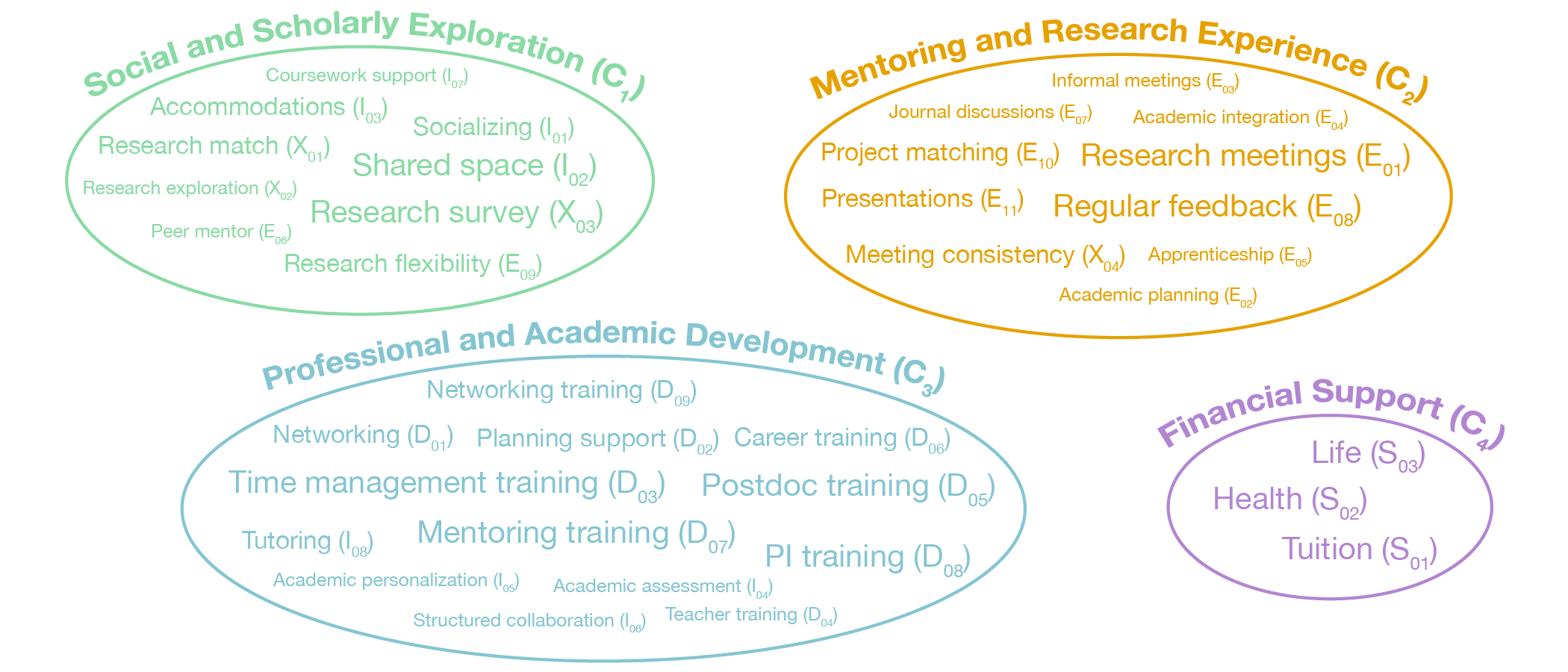}
    \caption{A visual representation of the four clusters in which each term represents a particular survey item.
    The colors correspond to those used to shade clusters in Fig.\ref{fig:ASES-net}: $C_1$ (green), $C_2$ (orange), $C_3$ (teal), or $C_4$ (purple). 
    The font size of each term reflects the relative importance of that survey item to the respective cluster, as determined by the hierarchical clustering shown in Fig.~\ref{fig:ASES-dendro}.
    Complete definitions of the items are presented in Table~\ref{tab:tab-comm}.}
    \label{fig:net-words}
\end{figure*}
\twocolumngrid

\noindent titions do overlap.
The one difference between $C_2$ and the PCA component $E$ is that two items from $E$ are not found in $C_2$ while one previously unloaded item is now included.
The additional aspect is $X_{04}$---``My research group meets at least once per week"---which is tied to the theme of research experiences.
The similarity shown here validates the coherence of the clusters and their relation to underlying themes.

Through investigating the hierarchical structure shown in the dendrogram (Fig.~\ref{fig:ASES-dendro}), we can further understand how the larger thematic constructs are built into the network.
In the ASES network, two types of clusters are observed.
The first type are clusters consisting of a set of core nodes and a few auxiliary ones (see $C_3$ in Fig.~\ref{fig:ASES-dendro}).
The second type includes clusters made up of two (or more) well-connected subclusters (see $C_1$ and $C_2$ in Fig.~\ref{fig:ASES-dendro}).
Qualitative interpretation allows for better understanding of what each distinct structure means for the larger themes.

In $C_3$, the core nodes indicate what is at the center of the theme that is represented by the cluster.
Many of the aspects at the core involve opportunities for the building of skills related to time management and networking.
While the whole cluster represents both academic and professional development, the central experiences that make up this cluster are related more to professional development.
This is shown by the relative text size in Fig.~\ref{fig:net-words}.

In $C_1$, one sub-cluster is made up of $I_{01}$, $I_{02}$, $X_{03}$, and $E_{09}$---all but one related to structures that make space for interaction (social activities, common areas, and seminars).
The other sub-cluster contains elements both of social support and scholarly exploration.
The sub-clusters are not necessarily meaningful by themselves and that the top level partition identified quantitatively through modularity optimization is the most meaningful partition of these aspects.

In $C_2$, one subgroup (consisting of $X_{04}$, $E_{07}$, $E_{10}$, and $E_{11}$) pertains to research group experiences, while the second sub-cluster (made up of $E_{01}$, $E_{02}$, $E_{03}$, $E_{04}$, $E_{05}$, and $E_{08}$) relates to interactions with research mentors.
The cohesion of the types of aspects observed in these sub-clusters can help interpret the complexity of the larger construct.
In $C_2$, the theme of mentoring and research experience is split between interaction with research groups and interaction with research mentors.

\subsection{Item-oriented interpretation}
We can also look at each item individually and evaluate its importance in the network.
In our analysis, we have done this through the weighted degree centrality.
Since we are looking at the backbone of the full survey item network, many nodes are only connected to the network by a few edges---two or three in most cases---resulting in relatively lower weighted degree measures (Table \ref{tab:tab-deg}).
However, there are a number of hubs within the backbone survey item network that have a high weighted degree, such as $D_{05}$: ``I attend activities for graduate students that include trainings or professional development on the role of a postdoc,'' $E_{10}$: ``The research project I am working on matches my research interests'', and $X_{01}$: ``I had or have support and flexibility from my department in finding my research interests.''
This indicates a type of decentralized network in which there are multiple high degree nodes, rather than a single hub (centralized network) or uniform degree measures across all nodes (distributed network).

Each of the largest three hub nodes within the decentralized network belongs to a different cluster.
We can better understand how these items function in the survey by pairing the high centrality measure with other information from network analysis.
For example, $D_{05}$ is connected to other nodes via negative temperature edges, which means that when other aspects of support structures (e.g., $D_{07}$ or $D_{08}$) are not experienced, $D_{05}$ is also absent from the student experience.
The nodes $E_{10}$ and $X_{01}$ not only are hubs within the network, but they are also connected to each other and create an edge that spans between two different clusters.
The similarity between these two aspects indicates that the fulfillment of personal research interest is important to the positive student experience of support structures.

The qualitative interpretation of ASES allows for better understating of the network analysis results.
This, in turn, enables diving deeper into the thematic constructs represented by the clusters to understand how individual aspects' contribute to the survey structure.
This approach is very promising for better understanding how respondents collectively view the experiences that are represented by the survey items.

\section{Conclusions}\label{sec:conclusion}
In this paper, we set out with four overall goals in mind.
First, to develop a methodology for creating a network from Likert-style survey responses.
Second, to understand how modular analysis of the network compares to PCA.
Third, to show how NA techniques can help validate the survey and identify redundant survey questions.
And fourth, to showcase the types of analysis that can be conducted through applying the NA derived thematic categories.

As part of NALS, our proposed method for generating a network involves the following steps:
\begin{itemize}[nosep]
    \item create bipartite network of respondents and response selections,
    \item project the network onto response selections using the edge weights to indicate number of respondents selecting both responses, 
    \item build an item relation matrix for each possible item pair, 
    \item calculate a similarity value between items and record the ratio of mutual agree to mutual disagree selections in temperature, and
    \item determine the backbone of the resulting network through identifying the most significant edges by similarity.
\end{itemize}
These steps are generalizable for any set of Likert-style survey responses, making the creation of networks based on many surveys used in PER possible.
For surveys that evaluate experiences or beliefs of closely connected and mutually reinforcing features (e.g., graduate program support structures or attitudes about learning science) a network representation is able to model how respondents relate each of these features to each other, building a hierarchical structure of thematic clusters and quantifying important items in the survey.

We found that the clusters identified through modular analysis and the previously calculated principal components had a strong resemblance.
While there are similarities between the two approaches for grouping survey items, we do see important differences in the partitions.
These differences have implications for the qualitative interpretation of the larger concepts that each aspect contributes to.
Additionally, an analysis of centrality measures of individual nodes identifies items that are more important for the network and cluster structure, giving additional information that is not easily assessed in PCA.
Though not explicitly explored in this work, this capability can be particularly valuable for comparing between distinct groups of respondents or identifying correlations between survey items and other factors.

We have also proposed two possible heuristics for identifying redundant survey items that aim to perform a similar function to that the covariance factors in PCA.
The first is based on the weighted degree centrality of each node, which can be used to prune the network of survey items that are seen as not strongly correlated to other items.
Alternatively, these items may be kept as they gather more unique responses, and the items of high centrality be pruned.
The second heuristic focuses on the hierarchical cluster structure of the network, and is useful for limiting the survey to only include items that are at the core of each cluster.

We have learned that the ASES as a whole is well designed with items clearly capturing larger themes of student experiences.
We have also found that the thematic clusters are slightly different than what was found in PCA, and are more complicated than previously identified.
For example, the NA revealed that certain items loaded into PCA components, such as $I_{04}$ and $I_{05}$ or $E_{07}$, are not central to the respective themes.
In contrast, items which were previously not loaded into any PCA component, such as $X_{01}$ and $X_{03}$, play important roles in the network structure, as confirmed by their centrality measure and placement in the network.
These items should be reconsidered and included in the survey distribution.

Overall, we have successfully developed a methodology by which NA can be used to investigate Likert-style surveys: NALS.
We have demonstrated NALS through the use of responses to the ASES survey, and compared it to a similar and widely used methodology, PCA.
A code repository has been created to facilitate use of NALS by other researchers~\cite{na-likert}.
The novelty of this approach brings a new way to understand how individuals collectively view interrelated aspects of complex phenomena, opening new avenues for investigations in PER and related fields.

\subsection{Limitations and future work}
We have made many decisions in developing our approach to creating the network and the analysis of that network.
Particular decisions have implications for how the methodology unfolds, thus it is important to understand the limitations of our approach.
In addressing these limitations, we also propose various threads for future research based on the affordances of NALS.

The final step in creating the network involves identifying the backbone of the full survey item network.
In this paper, we use the LANS algorithm to do so, however there are many choices of backbone algorithms and this choice may effect the resulting network.
We urge the pursuit of future investigations into the effect of different backbone algorithms in order to understand how they influence the interpretation of the networks.

When analyzing the ASES network, we use the weighted degree as our sole centrality measure.
Future work should be conducted to understand how more complex centrality measures can be used to aid interpretation of the Likert-style survey networks.
We also chose a single method of cluster detection, the fast-greedy algorithm.
For small networks, such as the ASES network, algorithms that optimize for modularity will always find the same partition.
For larger surveys that include many more items, it is important to understand the results of different algorithms and take these different partitions into consideration.

Although the approach is widely applicable, we have only demonstrated it for a single survey.
The specific features of the network we have investigated and the meaningful results may not be found in every instance.
We encourage other researchers to implement NALS on other Likert-style surveys in order to better understand the bounds of our technique.

\begin{acknowledgments}
This material is based upon work supported by the National Science Foundation (NSF) Graduate Research Fellowship Program under Grant No. DGE 1840340.
The views and conclusions contained in this paper are those of the authors and should not be interpreted as representing the official policies, either expressed or implied, of the NSF, or the U.S. Government.
\end{acknowledgments}

\appendix
\section{Item content by clusters}\label{app:clustering}
Table~\ref{tab:tab-comm} provides additional context for each cluster identified through NA.
It includes the code for each survey item along with the exact text that is used in the ASES instrument.
The items are grouped by cluster with the thematic cluster titles at the beginning of each grouping.
The principle component that each item was loaded into via PCA is provided, with numbers corresponding to those used by Sachmpazidi and Henderson~\cite{Sachmpazidi21-DSS}.
\vspace{5pt}

\onecolumngrid

\renewcommand{\arraystretch}{1.1}
\renewcommand{\tabcolsep}{6pt}
\begin{longtable}[H]{P{0.04\linewidth}p{0.85\linewidth}P{0.04\linewidth}}
\caption{
Cluster membership from the fast-greedy cluster algorithm. 
The first column indicates the question code while the second gives the item description, along with a shorthand name.
The component assignment, PC, is given in the last column. }
\label{tab:tab-comm} \\ 
\hline\hline
Code & Item content & PC \\ [1.2ex]
\endfirsthead
\caption[]{\emph{(Continued)}} \\
\hline\hline
Code & Item content & PC \\ [1.2ex]
\hline
\endhead
\hline
\multicolumn{3}{r}{\emph{(Table continued)}}
\endfoot
\endlastfoot
\hline 
    & {\bf $C_1$: Social and scholarly exploration support} & \\ \cline{2-2}
    $I_{01}$ & {\it Socializing}: The department hosts social activities (e.g., a welcome dinner, regular lunches) that are valuable in allowing me opportunities to share my thoughts and struggles with my peers, and discuss research areas. & 3 \\
    $I_{02}$ & {\it Shared space}: The department offered a space where students can build an academic and social community (e.g., student offices, rooms for tutoring, rooms for student leader organizations). & 3 \\
    $I_{03}$ & {\it Accommodations}: People in my department were supportive and caring about my accommodation needs when I first moved in to town. & 3 \\
    $E_{06}$ & {\it Peer mentor}: I have or had a senior peer mentor that provided invaluable resources and inducted me into departmental and/or laboratory cultures. & 1 \\
    $X_{01}$ & {\it Research match}: I had or have support and flexibility from my department in finding my research interests. & N/A \\
    $X_{02}$ & {\it Research exploration}: I had or have the opportunity to rotate through different research labs without making a commitment in order to find my research match. & N/A \\
    $X_{03}$ & {\it Research survey}: I attend(ed) a research seminar surveying the areas of expertise within the department. & N/A \\
    $E_{09}$ & {\it Research flexibility}: My research mentor was very flexible with my research assignments when I was struggling with one or more courses. & 1 \\
    $I_{07}$ & {\it Coursework support}: Whenever I face(d) a challenge succeeding on coursework, someone from my department helped me overcome it. & 3 \\  \cline{2-2} 
    & {\bf $C_2$: Mentoring and research experience}\\\cline{2-2}
    $E_{01}$ & {\it Research meetings}: I have frequent meetings with my mentor to discuss on my research progress and any challenges I face. & 1 \\
    $E_{02}$ & {\it Academic planning}: My mentor(s) helped me selecting courses and develop my academic plans. & 1 \\
    $E_{03}$ & {\it Informal meetings}: I have informal meetings with my mentor(s) where I get assistance or support with any issues I face (for example, on issues such as life-work balance, develop social network, set future goals, access health care resources, etc.). & 1 \\
    $E_{04}$ & {\it Academic integration}: My mentor(s) helped me integrate into the program and the physics community. & 1 \\
    $E_{05}$ & {\it Apprenticeship}: My mentor(s) taught me what it means to be a research physicist and a scholar. & 1 \\
    $X_{04}$ & {\it Meetings consistency}: My research group meets at least once per week. & N/A \\
    $E_{07}$ & {\it Journal discussions}: In my research group meetings, we devote time in reading and discussing about the current state of knowledge in the field. & 1 \\
    $E_{08}$ & {\it Regular feedback}: I have regular meetings with my research mentor and receive feedback on a regular basis. & 1 \\
    $E_{10}$ & {\it Project matching}: The research project I am working on matches my research interests. & 1 \\
    $E_{11}$ & {\it Presentations}: I have presented or am planning to present my research at a group meeting or in a journal club. & 1 \\ \cline{2-2}
    & {\bf $C_3$: Professional and academic development} \\ \cline{2-2}
    $I_{04}$ & {\it Academic assessment}: In the beginning of my program, I took a precourse assessment that was designed to measure my incoming preparation. & 3 \\
    $I_{05}$ & {\it Academic personalization}: I was offered a personalized coursework plan in my graduate program.	& 3 \\
    $I_{06}$ & {\it Structured collaboration}: The faculty, postdocs or experienced TAs lead guided group-work sessions to encourage students work collaboratively on concepts covered in core courses. & 3 \\
    $D_{01}$ & {\it Networking}: I attend mini-conferences where students from nearby universities can share research progress and learn networking skills. & 2 \\
    $D_{02}$ & {\it Planning support}: At the beginning of each semester, my faculty advisor(s) and I developed time-management plan that help me identify areas where my time could be used more effectively. & 2 \\
    $D_{03}$ & {\it Time management training}: My department hosts a seminar that focuses on time management skills. & 2 \\
    $I_{08}$ & {\it Tutoring}: My department makes tutoring available to graduate students. & 3 \\
    $D_{04}$ & {\it Teaching training}: I attend activities for graduate students that include trainings or professional development on best practices for effective teaching. & 2 \\
    $D_{05}$ & {\it Postdoc training}: I attend activities for graduate students that include trainings or professional development on the role of a postdoc. & 2 \\
    $D_{06}$ & {\it Career training}: I attend trainings that focus on how to maximize my chances of finding a career that is a good fit for my interests and skills. & 2 \\
    $D_{07}$ & {\it Mentoring training}: I attend training on learning about mentoring skills as future faculty or postdoc. & 2 \\
    $D_{08}$ & {\it PI training}: I attend training on organizing a research laboratory. & 2 \\
    $D_{09}$ & {\it Networking training}: I attend activities where I can learn about effective networking. & 2 \\ \cline{2-2}
    & \bf{ $C_4$: Financial support} \\ \cline{2-2}
    $S_{01}$ & {\it Tuition}: My tuition is covered for my entire program. & 4 \\
    $S_{02}$ & {\it Health}: My college, department, or program offers me health benefits. & 4 \\
    $S_{03}$ & {\it Life}: I have no financial concerns about completing my degree. & 4 \\
    \hline\hline
\end{longtable}

\twocolumngrid

\pagebreak
\newpage
%

\end{document}